\documentclass[twocolumn]{emulateapj}

\usepackage{mathrsfs}
\usepackage{graphicx} %for EPS, PS files

\newcommand{\sgr}{Sgr\,A*}
\newcommand{\msun}{M_\odot}
\newcommand{\mfrac}[2]{\mbox{$\frac{#1}{#2}$}}

\shorttitle{Probing \sgr\ with a pulsar} \shortauthors{Liu et al.}

\begin{document}

\title{Prospects for Probing the Spacetime of \sgr\ with Pulsars}

\author{
K.~Liu$^{1,2}$, N.~Wex$^1$, M.~Kramer$^{1,2}$, J.~M.~Cordes$^{3}$,
and T.~J.~W.~Lazio$^{4}$ }

\affil{$^{1}$Max-Planck-Institut f\"ur Radioastronomie, Auf dem H\"ugel 69,
             53121, Bonn, Germany}
\affil{$^{2}$University of Manchester, Jodrell Bank Centre for
             Astrophysics, Alan-Turing Building, Manchester M13 9PL, UK}
\affil{$^{3}$Astronomy Dept., Cornell Univ., Ithaca, NY 14853, USA}
\affil{$^{4}$Jet Propulsion Laboratory, M/S 138-308, 4800 Oak Grove Dr.,
             Pasadena, CA 91109, USA}

\begin{abstract}
The discovery of radio pulsars in compact orbits around \sgr\ would
allow an unprecedented and detailed investigation of the spacetime
of the supermassive black hole. This paper shows that pulsar timing,
including that of a single pulsar, has the potential to provide
novel tests of general relativity, in particular its cosmic
censorship conjecture and no-hair theorem for rotating black holes.
These experiments can be performed by timing observations with
100\,$\mu$s precision, achievable with the Square Kilometre Array
for a normal pulsar at frequency above 15\,GHz. Based on the
standard pulsar timing technique, we develop a method that allows
the determination of the mass, spin, and quadrupole moment of \sgr,
and provides a consistent covariance analysis of the measurement
errors. Furthermore, we test this method in detailed mock data
simulations. It seems likely that only for orbital periods below
$\sim 0.3$\,yr is there the possibility of having negligible
external perturbations. For such orbits we expect a $\sim 10^{-3}$
test of the frame dragging and a $\sim 10^{-2}$ test of the no-hair
theorem within 5 years,  if \sgr\ is spinning rapidly. Our method is
also capable of identifying perturbations caused by distributed mass
around \sgr, thus providing high confidence in these gravity tests.
Our analysis is not affected by uncertainties in our knowledge of
the distance to the Galactic center, $R_0$. A combination of pulsar
timing with the astrometric results of stellar orbits would greatly
improve the measurement precision of $R_0$.
\end{abstract}


\maketitle

%%%%%%%%%%%%%%%%%%%%%%%%%%%%%%%%%%%%%%%%%%%%%%%%%%%%%%%%%%%%%%%%%%%%%%%%%%%%%%%%

\section{Introduction} \label{sec:intro}

One of the most intriguing results of general relativity (GR) is the uniqueness
theorem for the stationary black hole solutions of the Einstein-Maxwell
equations \citep[see][and references therein]{heu98}. This uniqueness theorem
states that (under certain conditions) all stationary
electrovac\footnote{Electrovac spacetimes are the solutions of the
Einstein-Maxwell equations.} black hole spacetimes with a non-degenerate horizon
are described by the Kerr-Newman metric. It implies that in GR all stationary
black holes are parameterized by only three parameters: mass ($M$), spin ($S$)
and electric charge (``black holes have no hair''). All uncharged black hole
solutions are described by the Kerr metric and, therefore, uniquely determined
by $M$ and $S$. Astrophysical black holes are believed to be the result of a
gravitational collapse. During this collapse all the properties of the
progenitor, apart from mass and spin, are radiated away by gravitational
radiation while the gravitational field asymptotically approaches its stationary
configuration \citep{pri72a,pri72b}. The outer spacetime of an astrophysical
black hole should therefore be described by the Kerr metric.\footnote{Strictly
speaking, this is only true for a certain approximation since, to some
extent, astrophysical black holes will be influenced by nearby masses
(accretion, orbiting objects). We will address this issue for \sgr\
in this paper at the end of the discussion Section.} Since the Kerr metric has
a maximum spin at which it still exhibits an event horizon, Penrose's cosmic
censorship conjecture (CCC) within GR \citep{pen79} requires the dimensionless
spin parameter $\chi$ to satisfy
\begin{equation}
  \chi \; \equiv \; \frac{c}{G}\,\frac{S}{M^2} \; \le 1 \;
\end{equation}
A measured value for $\chi$ that exceeds 1 would pose a serious problem for our
understanding of spacetime, since this would indicate that either GR is wrong or
that a region may be visible to the outside universe, where our present
understanding of gravity and spacetime breaks down.

As a result of the no-hair theorem, all higher multipole moments ($l \ge 2$) of
the gravitational field of an astrophysical black hole can be expressed as a
function of $M$ and $S$ \citep{han74}. In particular, the dimensionless
quadrupole moment $q$ satisfies the relation
\begin{equation}
  q \; \equiv \; \frac{c^4}{G^2}\,\frac{Q}{M^3} \; = \; -\chi^2 \;.
\end{equation}
A measurement of the quadrupole moment, in combination with a mass and a spin
measurement, would therefore provide a test of the no-hair theorem for Kerr
black holes.

Some of the clearest evidence for the existence of black holes comes from the
monitoring of $\sim 30$ stellar orbits in the center of our Galaxy
\citep{sog+02,gsw+08,get+09}, where the shortest orbital period, $P_{\rm b}$, is
16 years. Known by its radio nomenclature of \sgr, current estimates put the
mass of this black hole to around $4 \times 10^6\,\msun$. A black hole of that
size at a distance of 8\,kpc is an ideal laboratory for black-hole physics,
strong-field gravity and, in particular, a test of the no-hair theorem for Kerr
black holes \citep[e.g.][]{psa08,pj10}. It has been shown by \cite{wil08} that
the discovery of stars in highly eccentric ($e\sim 0.9$) orbits very close to
\sgr\ ($P_{\rm b} \la 0.1\,{\rm yr}$) could be used to test the general
relativistic no-hair theorem. This experiment requires an astrometric precision
at the level of 10\,$\mu$as, which seems achievable with the upcoming
infrared-astrometry experiments, such as GRAVITY \citep{epb+09}. At a distance
of 8\,kpc, an angle of 10\,$\mu$as corresponds to a length scale of $\sim
10^7$\,km.

On the other hand, if close-in pulsars could be found and tracked in their
orbits, even for those with poor timing precision the time-of-arrival (TOA) for
a (integrated) pulsar signal can be measured with an uncertainty of a few
milliseconds, corresponding to a light-travel distance of $\sim 10^3$\,km.
Moreover, a phase-connected solution with an appropriate timing model allows a
determination of the pulsar orbit, which provides even more precision. Hence, as
shown by \cite{wk99} and \cite{kbc+04}, a pulsar in orbit around the
supermassive black hole in the Galactic center (GC) would provide an ideal probe
to measure the mass, the spin, and the quadrupole moment of \sgr, and
consequently test the no-hair theorem for Kerr black holes. In their discussion
on ``bumpy black holes'', \cite{vh10} suggest that a pulsar in orbit around a
black hole could be used for mapping its multipole moment structure. In a recent
publication \cite{asm10} have discussed the importance of frame dragging and the
quadrupole moment for stars and pulsars in orbit around \sgr, based on numerical
integration of geodesics in a Kerr spacetime. For pulsars, however, the results
in this paper are only indicative.\footnote{The \cite{asm10} results for a
pulsar are not derived from a consistent covariance analysis based on a timing
model that incorporates all the relevant effects simultaneously. Moreover,
pulsar timing is treated as a radial velocity measurement experiment, which is
incorrect. In fact, pulsar timing makes use of phase-connected solutions for the
rotational phase of a pulsar leading to a precision in the parameter estimations
that can be several orders of magnitude better
\citep{lk05}. Also, the timing precision assumed by \cite{asm10}
seems too optimistic for a GC pulsar, as will become clear from the analysis
presented in this paper.} \citet{wcp+09,wjc+09} have shown that if a pulsar is
found in a very close orbit around \sgr\ (e.g.~$P_{\rm b} \la 1\,{\rm day}$),
observers at the Earth can receive additional pulses travelling along a path
that is strongly bent by the gravitational potential of the black hole. The
exploitation of this information would provide unique constraints on the
strongly curved spacetime geometry near \sgr.

In this paper we will demonstrate how the mass, the spin (magnitude and
orientation) and the quadrupole moment of \sgr\ can be determined from timing a
pulsar in a sufficiently tight orbit around the supermassive black hole. Our
analysis is based on simulated TOAs and a timing model that allows for a
phase-connected solution, consistently accounting for the relativistic effects
in the motion of the pulsar and the propagation of the radio signals. Based on
this, we can determine the expected precision for the individual parameters
while accounting for all the correlations between the parameters. In all
discussions and simulations we focus on the leading order in the individual
effects of interest. We are well aware that in an actual timing model for orbits
close to \sgr\ we need to account for higher order effects in the orbital motion
and signal propagation. For many of the effects discussed below, higher order
terms have already been calculated \citep[e.g.][]{ds88, wex95, kop97b, kg05}.
However, to estimate the expected precision and covariances in the parameter
determination, it is sufficient to use a timing model that combines just the
leading terms of all contributions relevant here. The second order terms
contribute at a $\sim\beta_{\rm O}^2$ level, where $\beta_{\rm O}$ is the
orbital velocity parameter introduced by \cite{dt92}. For a test particle in
orbit around a mass $M$
\begin{eqnarray}
  \beta_{\rm O} &=& \left(\frac{2\pi GM}{c^3P_{\rm b}}\right)^{1/3} \nonumber \\
          &\simeq& 0.0158
                 \left(\frac{M}{4\times 10^6\,M_\odot}\right)^{1/3}
                 \left(\frac{P_{\rm b}}{1\,{\rm yr}}\right)^{-1/3}\;.
\end{eqnarray}
As an example, for an orbital period of 0.1\,yr we find $\beta_{\rm O}^2 \approx
10^{-3}$. At this point it is worth mentioning, that for the most relativistic
binary pulsar, the double pulsar, one finds $\beta_{\rm O}^2 \simeq 4.3\times
10^{-6}$ \citep{kw09}, which nicely illustrates how much more relativistic a
pulsar in a $P_{\rm b} \la 1\,{\rm yr}$ orbit around \sgr\ would be.

Previous studies suggest that about 1000 pulsars can be expected to
be orbiting \sgr\ with periods less than 100 years
\citep{cl97,gso+03,pl04}\footnote{The result from the GC survey by
\cite{mkf+10} indicated that the actual number of such pulsars may
be less.}, and some of them may be associated with remnants of the
observed S-star population in the neighborhood. Deep pulsar searches
towards the GC region were already conducted with a few radio
telescopes (Effelsberg, Green Bank, Parkes) at frequencies up to
15\,GHz \citep[e.g.][]{kkl+00,mkf+10}. Five pulsars were found no
more than 200\,pc away from \sgr\ \citep[][Kramer et al.~private
communication]{jkl+06,dcl09}, which is consistent with the estimated
large pulsar population within that region. However, they are not
close enough to the supermassive black hole to probe its
gravitational field. In this paper we will focus on timing
observations of such pulsars, and show how far they could take us in
probing the gravitational field of \sgr, provided the system is
found to be sufficiently free of external perturbations. The main
purpose therefore is the development of the methodology, and the
estimation of its potentials in testing the Kerr nature of \sgr\
based on mock data simulations. For further elaboration on either
the existence of pulsars in close orbits around \sgr\ or the search
for them we refer to the rich literature \citep{lk05} and future
work in progress. In Section~\ref{sec:timing} we discuss the
expected timing precision and the orbital periods required for our
measurements. Section~\ref{sec:mass} presents the various
relativistic effects that can be used to determine the mass of \sgr\
and, based on simulations, the expected precision in the mass
measurement. In Section~\ref{sec:spin} we show how the spin can be
extracted from the timing measurement, how this information can be
used to test the CCC, and how the presence of a distributed mass in
the vicinity of \sgr\ would affect this measurement.
Section~\ref{sec:quadrupole} provides the details on the quadrupole
measurement and the no-hair theorem test. In
Section~\ref{sec:discussion} we summarize and discuss our findings.

%%%%%%%%%%%%%%%%%%%%%%%%%%%%%%%%%%%%%%%%%%%%%%%%%%%%%%%%%%%%%%%%%%%%%%%%%%%%%%%%

\section{Timing a pulsar in orbit around \sgr}
\label{sec:timing}

Pulsar timing involves measuring the TOAs of a pulsar's pulses and monitoring
them on a timescale of years \citep[e.g.][]{tay92}. The precision of TOA
measurements of young pulsars near the GC, by future telescopes, will mainly be
limited by three effects that have significantly different dependencies on
observing frequency: firstly the signal-to-noise ratio of the measured pulses,
secondly the pulse phase jitter intrinsic to the pulsar, and thirdly the changes
in pulse shape caused by interstellar scintillation \citep{cs10,lvk+11}. The
first effect is independent of frequency under our assumption that the pulse
width does not change with frequency. In reality, the pulse width does evolve
but that is secondary to the overall timing error. The second and third effects
are strongly frequency dependent, due to the steep pulsar spectrum and the pulse
broadening caused by the large amount of scattering from the high electron
density in the ionized gas near the GC. The strong dependence of scattering on
frequency ($\propto f^{-4}$, see the next paragraph) implies that observations
need to be made at much higher frequencies than are typically used for pulsar
timing.

\begin{figure}[ht]
\begin{center}
\includegraphics[width=8.5cm]{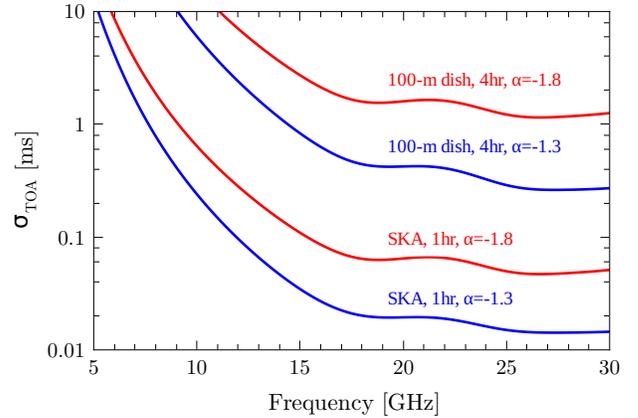}
\caption{
Predicted TOA measurement precision of a young pulsar near the GC for two
different spectral indices $\alpha$. The curves take into account pulse phase
jitter intrinsic to the pulsar, pulse broadening from scattering along the
entire line of sight, and from the finite number of scintles included in the
measurement. Scattering is dominated by a region of high plasma density that
surrounds the GC. We assumed a four hour integration time using a 100-meter
radio telescope and a one hour integration time using an SKA-like telescope,
both with a highest operating frequency of 30\,GHz and a bandwidth of 1~GHz. It
is found that observational frequencies above 15\,GHz favors pulsar timing
observation, where 100\,$\mu$s TOA precision seems achievable, in particular
with the SKA. The parameters used for this calculations can be found in the
text.
\label{fig:GCsigma}}
\end{center}
\end{figure}

There have been previous studies on optimizing the observational frequency for
the purpose of pulsar searches towards the central parsec region
\citep{cl97,mkf+10}. Fig.~\ref{fig:GCsigma} shows the estimated timing
precision for a canonical pulsar near \sgr\ as a function of the observing
frequency. The calculation of the achievable TOA uncertainty $\sigma_{\rm TOA}$
accounts for three contributions:
\begin{equation}
\sigma^2_{\rm TOA} = \sigma^2_{\rm rn} + \sigma^2_J + \sigma^2_{\rm scint} \;.
\end{equation}
Here $\sigma_{\rm rn}$, $\sigma_J$, and $\sigma_{\rm scint}$ represent the
uncertainties contributed by radiometer noise, pulse phase jitter and
interstellar scintillation, respectively, which can be calculated by following
e.g.~\citet{cs10}. Specifically, we use a spin period $P = 0.5$~s, an intrinsic
pulse width  $W_i = 10$~ms, and a period-averaged flux density $S_{1400} =
1$~mJy at 1.4\,GHz. For a 100-m diameter dish and the SKA, we use a gain of
2\,K~Jy$^{-1}$, 100\,K~Jy$^{-1}$, and an integration time of 4\,hr, 1\,hr,
respectively. Two different spectral indices of the pulsar flux density, which
typifies many of these measured for pulsars \citep{mkkw00a}, are used in our
calculations. The scattering time scale is estimated to be $\tau_{\rm scat}
\approx 2.3\times10^6$\,ms at 1\,GHz, as derived from the observed scattering
diameter of \sgr\ and the estimated location of the scattering material along
the line of sight, the latter as incorporated in the NE2001 model \citep[$\ell =
b = 0$ and $D$=8.5~kpc, ][]{cl02}. For this large amount of scattering, we use a
scaling of $\tau_{\rm scat}\propto f^{-4}$ \citep[e.g.][]{lkm+01} rather than
the often used Kolmogorov scaling $\tau_{\rm scat}\propto f^{-4.4}$
\citep[e.g.][]{ric90}, because the dominant length scale is less than the inner
scale of the wavenumber spectrum for the electron density. Note that all
potential pulsars with close orbits of interest for the GR tests will be seen
along essentially the same line of sight as \sgr, so one can assume that their
lines of sight will have the same scattering characteristics. The system
temperature (e.g. $\simeq40$\,K at 15\,GHz) is calculated by summing the radio
background, receiver temperature and emission of the atmosphere. It clearly
follows from Fig.~\ref{fig:GCsigma}, that with a radio telescope like SKA TOA
uncertainties of below 100\,$\mu$s seem likely for an observational frequency
above 15\,GHz, similar to the result of optimized searching frequency. A
detection of MSPs in the Galactic centre is unlikely \citep{cl97,mkf+10}, so
they are not considered in the following simulations. However, we will show that
the black hole properties can already be extracted by finding and timing a
relatively slow pulsar. If a MSP were to be found after all, the experiment may
reach a correspondingly higher precision.

Precision of long-term timing of a young pulsar is often limited by
irregularities of the pulsar's rotation, supposedly associated with either the
internal super-fluid flux \citep[e.g.][]{mw09}, or external magnetic field
activity \citep[e.g.][]{lhk+10}. The amplitude of the low-frequency noise
resulting from these instabilities varies from about 10\,$\mu$s to 100\,ms
\citep{hlk10}, and in some cases the noise can be modeled by following the
approach proposed by \cite{lhk+10} to improve the timing precision by order of
magnitudes. Consequently, a TOA precision of 100\,$\mu$s is a reasonable
fiducial value, which we will use in our simulations below.

\begin{figure}[ht]
\begin{center}
\includegraphics[width=8.5cm]{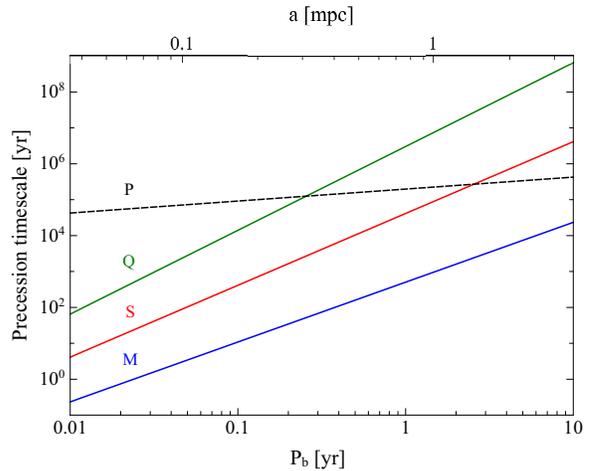}
\caption{Timescales of secular orbital precession for a pulsar in
orbit around \sgr\ as a function of orbital period $P_{\rm b}$
(semi-major axis $a$). The letters M, S, Q and P stand for the
contribution by the mass monopole (pericenter advance), the spin
(frame dragging), the quadrupole moment, and stellar perturbation,
respectively. Here we assume an orbital eccentricity of 0.5 and
$10^3$ one solar mass objects within 1\,mpc around \sgr. As a
comparison, the Schwarzschild radius of \sgr\ is $\sim 4\times
10^{-4}$\,mpc. \label{fig:precession_rates}}
\end{center}
\end{figure}

Although the purpose of this paper is to discuss potential gravity tests
with a pulsar in orbit around \sgr, provided the system is found to be
sufficiently clean, we nevertheless would like to complete this Section with a
brief discussion on possible effects that could complicate or even spoil these
tests. \cite{mamw10} and \cite{sw11} have shown that for orbits with an orbital
period $P_{\rm b}$ larger than 0.1\,yr, it becomes likely that the distribution
of stars in the vicinity causes ``external'' perturbations of the orbital motion
of the pulsar and prevent a clean test of the no-hair theorem or even a
measurement of the Lense-Thirring effect. In order to evaluate the significance
of the perturbation, following the analysis of \cite{mamw10}, in
Fig.~\ref{fig:precession_rates} the relation of precessional timescale against
orbital size is presented for four different contributions: the pericenter
advance, the frame dragging effect, the black-hole quadrupole, and the
surrounding mass distribution. Here we assume $10^3$ \citep[the highest
number applied in][]{mamw10} one solar mass objects isotropically distributed
within 1\,mpc around \sgr. Similar to \cite{mamw10} we do not consider the
influence of objects outside the central 1\,mpc region. One can see that for
wide orbits ($P_{\rm b} \sim 10\,{\rm yr}$) the pericenter advance is still
significantly larger than the precession caused by external perturbations. This
suggests that for orbital periods less than about 10 years the measured
$\dot{\omega}$ can be used to tightly constrain the black hole mass. The frame
dragging will be dominant over the stellar noise if the orbital period is less
than 0.5 years, while only for orbital periods $\la 0.1$\,yr the (secular)
contribution of the quadrupole moment is expected to be significantly above the
external perturbation. We note that the assumptions applied to calculate
the precessional timescale by stellar perturbation may not be secure, as the
actual stellar components and mass distributions within the central pc
(especially the central mpc) are still not fully understood. For instance, the
precessional torque could be larger if there exists a high fraction of massive
objects near \sgr\ due to mass segregation \citep{okl09,kha09,kt11}, or a
significant anisotropy in the distribution of the surrounding masses. In fact,
it has been argued by \cite{mamw10} that a high fraction of $10\,M_\odot$ black
holes in this region would make astrometric tests of general relativity
problematic at all radii. Concerning pulsars, however, as will be discussed in
Sections~\ref{ssec:perturb_spin} and \ref{ssec:q_simu}, the gravitomagnetic and
quadrupolar field of \sgr\ will result in unique features in the timing
residuals, which can be tracked well with high precision timing observations,
and one can still expect to be able to extract the \sgr\ spin and quadrupole
moment from the timing data, to some extent. This however depends strongly on
the details of the external perturbations, which only will be known once a
pulsar is discovered in that region.

%%%%%%%%%%%%%%%%%%%%%%%%%%%%%%%%%%%%%%%%%%%%%%%%%%%%%%%%%%%%%%%%%%%%%%%%%%%%%%%%

\section{Mass measurement} \label{sec:mass}

The current best estimate for the mass of \sgr\ gives $4.30 \pm 0.20_{(\rm
stat)} \pm 0.30_{(\rm sys)} \times 10^6\,\msun$ \citep{get+09}\footnote{The main
uncertainty is from the limited knowledge of the distance to the GC.}. The
proposed method has the potential to improve the measurement accuracy by a
factor of $\sim 10^5$. This is possible pulsar timing. Indeed, the most precise
mass measurements for stars (other than the Sun) come from pulsar timing
observations \citep{jhb+05,fbw+11}. Those are achieved in binary pulsar systems,
where in addition to the Keplerian parameters one can determine a set of
post-Keplerian (PK) parameters as theory independent relativistic corrections.
In any theory of gravity the PK parameters are functions of the two a priori
unknown masses of the system, which can be determined once two PK parameters
have been obtained \citep[see][for definitions of the Keplerian and PK
parameters]{dt92}. Since in our case the mass of the pulsar can be neglected in
comparison to the mass of the black hole, in general one PK parameter is
sufficient to estimate the mass of \sgr\ with a precision at the $10^{-6}$
level. In the following we briefly discuss three relativistic effects that can
be used for a mass determination, and present the results of extensive
simulations at the end of this Section. As mentioned in Section~\ref{sec:intro},
in a discussion of measurement precision it is sufficient to keep the first
order terms in the description of these effects.

%%%%%%%%%%%%%%%%%%%%%%%%%%%%%%%%%%%%%%%%%%%%%%%%%%

\subsection{Post-Keplerian parameters and mass determination}

In eccentric binary pulsars the {\em precession of periastron}, $\dot\omega$, is
usually the first PK parameter that can be measured with high precision. It
allows the determination of the total mass of the system, which in our case can
be equaled with the mass of the black hole $M_{\rm BH}$. To first order one
finds \citep{rob38}:
\begin{eqnarray}
  \dot\omega
    &\simeq& \frac{3}{1 - e^2} \left(\frac{2\pi}{P_{\rm b}}\right)^{5/3}
             \left(\frac{GM_{\rm BH}}{c^3}\right)^{2/3} \nonumber\\
    &\simeq& (0.269^{\circ}{\rm /yr})\,\frac{1}{1 - e^2}
             \left(\frac{P_{\rm b}}{1\,{\rm yr}}\right)^{-5/3}
             \left(\frac{M_{\rm BH}}{4\times 10^6\,\msun}\right)^{2/3}
             \;,\label{eq:omegadot} \nonumber \\
             &&
\end{eqnarray}
where $e$ is the orbital eccentricity. As an example, for a 0.3\,yr orbit with
an eccentricity of 0.5 the orbit precesses with a rate of about 2.7 degrees per
year. After five years of weekly observations with a timing uncertainty of
$100\,\mu$s, this precession will be measured with a fractional precision of
better than $10^{-7}$. This, however, is not the precision with which the mass
of \sgr\ can be determined. If the black hole is rotating, a significant
fraction of the pericenter precession can come from the frame dragging
\citep{bo75}. Depending on the orientation and the spin of the black hole, this
could be up to about 1\% of the total precession. A measurement of the spin and
the orientation of the black hole would allow to correct for this Lense-Thirring
contribution $\dot\omega_{\rm LT}$. But as we will show later, we will use the
observed $\dot\omega$ and the mass measurement from other relativistic effects
to calculate $\dot\omega_{\rm LT}$ and use it in the spin determination
($\dot\omega \equiv \dot\omega_{\rm M} + \dot\omega_{\rm LT}$).

The {\em Einstein delay} is a combination of the second order Doppler effect and
gravitational redshift. From its amplitude $\gamma_{\rm E}$, which is a PK
parameter, one can determine the mass of the black hole, since (to first
post-Newtonian order)
\citep{bt76}
\begin{eqnarray}
   \gamma_{\rm E}
     &\simeq& 2 e \left( \frac{P_{\rm b}}{2\pi}  \right)^{1/3}
                  \left( \frac{GM_{\rm BH}}{c^3} \right)^{2/3}
     \nonumber\\
     &\simeq& (2500\,{\rm s})\,e
             \left(\frac{P_{\rm b}}{1\,{\rm yr}}\right)^{1/3}
             \left(\frac{M_{\rm BH}}{4\times 10^6\,\msun}\right)^{2/3}
             \;.
\end{eqnarray}
For a 0.3 yr orbit with an eccentricity of 0.5 the amplitude of the Einstein
delay will be of order 800 seconds. However, the Einstein delay is a priori not
separable from the Roemer delay\footnote{The Roemer delay is defined as
$\Delta_{\rm R}=-\hat{\bf K}_0\cdot{\bf r}$, where $\hat{\bf K}_0$ is the unit
vector of the line-of-sight and {\bf r} is the position vector of the pulsar
with respect to the barycenter of the binary system. The Roemer delay describes
the contribution of the pulsar motion to the time delay.}, and can only be
measured with sufficient accuracy after some time, when the relativistic
precession has changed the orbital orientation sufficiently. For a pulsar in an
0.3\,yr orbit this is already the case after a few orbits. After a few years the
Einstein delay can be measured with high precision, as will be shown in the
simulations below. The dragging of inertial frames in the vicinity of the black
hole also affects the Einstein delay. However, this occurs only at higher orders
\citep{wex95}, which in most cases can be neglected or easily accounted for in a
combined mass and spin measurement.

The {\em Shapiro delay} accounts for the extra light travelling time due to the
curvature of space-time caused by the existence of surrounding masses (here
mainly \sgr). The Shapiro delay contains two separately measurable PK
parameters, the mass of the black hole $M_{\rm BH}$ and $\sin i$. The signal is
usually only sufficiently strong for edge-on systems \citep[e.g.][]{ksm+06}, but
in our case even for a face-on orbit ($i = 0$) the effect will be significant
due to the large mass of \sgr, if the orbit is eccentric. Using the equation of
\cite{bt76},
\begin{eqnarray}
  \Delta_{\rm S}
    &\simeq& \frac{2GM_{\rm BH}}{c^3}\,
             \ln \left( \frac{1 + e \cos\varphi}
                             {1 - \sin i\sin(\omega + \varphi)} \right)
    \nonumber\\
    &\simeq& (39.4\,{\rm s})
             \left(\frac{M_{\rm BH}}{4\times 10^6\,\msun}\right) \,
             \ln \left(\frac{1 + e \cos\varphi}
                            {1 - \sin i\sin(\omega + \varphi)} \right)
                            \;, \label{eq:sdelay} \nonumber \\
                            &&
\end{eqnarray}
as a first order estimation, one can see that for an eccentricity of 0.5 the
Shapiro delay for $i = 0$ amounts to about 40 seconds. This already indicates
that the Shapiro delay allows a precise mass determination, even for a pulsar
with poor timing precision. Apart from containing $M_{\rm BH}$ directly,
the Shapiro delay gives a second, though indirect, access to the \sgr\ mass via
$\sin i$ and the mass function. One finds
\begin{equation} \label{eq:mf}
  GM_{\rm BH} \simeq  \left( \frac{cx}{\sin i} \right)^3
                      \left( \frac{2\pi}{P_{\rm b}} \right)^2 \;,
\end{equation}
where $x$ is the projected semi-major axis of the pulsar orbit (in light
seconds), which is an observable Keplerian parameter. It depends on the orbital
eccentricity and inclination, which of the two is more constraining.

In addition, there are significant contributions to the signal propagation
caused by frame dragging. A first order analytic equation for this effect can be
found in \cite{wk99}. From this it is clear that the frame dragging can have a
significant contribution to the propagation delay, but in most cases will have a
distinct signature that can be fitted for, leading at the same time to a precise
mass measurement and a lower limit on the spin parameter $\chi$. Contributions
from higher order multipole moments and light bending effects can easily be
accounted for in an analytic way \citep[see e.g.][]{kop97b}.

The inclination of the pulsar orbit with respect to the line-of-sight $i$
(modulo a $\pi - i$ ambiguity, see Fig.~\ref{fig:angles}) can be obtained either
directly from the Shapiro delay, as explained above, or via Eq.~(\ref{eq:mf}) by
using the mass, $M_{\rm BH}$, derived from any other PK parameter. Therefore, in
Section~\ref{sec:spin} and \ref{sec:quadrupole} where the determination of spin
and quadrupole is presented, we can treat the inclination angle as a parameter
that is known with sufficient precision. A brief discussion on the $\pi - i$
ambiguity can be found in Section~\ref{ssec:spin deter}.

%%%%%%%%%%%%%%%%%%%%%%%%%%%%%%%

\subsection{Simulations} \label{ssec:mass_simu}

The simulations performed in this paper mainly contain two steps: creating
TOAs and determining parameters as well as their measurement uncertainties.
Firstly, the TOAs are created regularly regarding to solar system barycentric
time and then combined with the three time delays (Roemer, Einstein and Shapiro,
see the above Subsection) to account for the changes in the signal arrival time
due to the orbital motion of the pulsar around \sgr. Next the simulated TOAs are
passed to the TEMPO software package. Based on a timing model, TEMPO performs a
least-square fit to yield a phase-connected solution of the TOAs, and determines
the model parameters. The measurement uncertainties of these parameters are
calculated via a covariance matrix. This is the standard procedure of pulsar
timing observations, and explained in great detail in \cite{tay94b,lk05,
hem06,ehm06}. Most of the timing models used in this paper are part of the TEMPO
standard implementation available as a download from the sources given in these
references. When ever we use an extension to these well tested models, to
account for specific effects, which are not covered by the standard software, we
will mention this explicitly in the corresponding Section.

\begin{figure}[ht]
\begin{center}
\includegraphics[width=8.5cm]{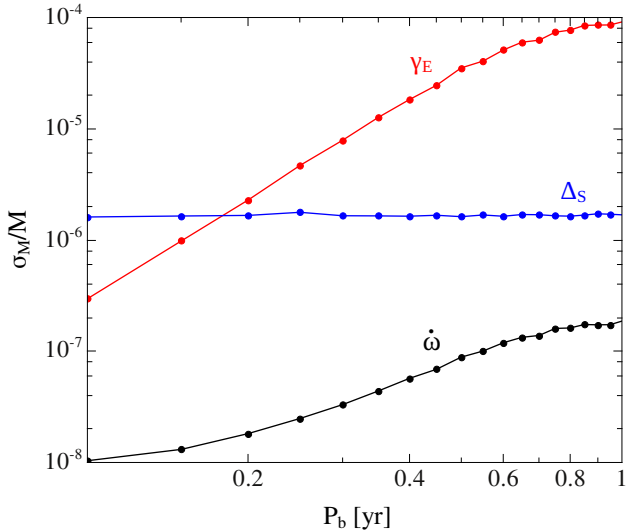}
\caption{Simulated fractional precision for the mass determination of \sgr\ as a
function of the orbital period $P_{\rm b}$, obtained from three different
relativistic effects: precession of the orbit ($\dot\omega$), Einstein delay
($\gamma_{\rm E}$), and Shapiro delay ($\Delta_{\rm S}$). The mass
determinations are based on simulated data, assuming weekly TOAs with an
uncertainty of 100\,$\mu$s over a time span of five years. We used an orbital
eccentricity $e$ of 0.5 and an orbital inclination $i$, relevant for the Shapiro
delay, of $60^\circ$. The simulations were done for a non-rotating black hole.
Note that for various practical reasons (such as the uncertainty in the pulsar
mass), a precision below $10^{-7}$ seems unrealistic. Also, as explained in the
text, for a rotating black hole $\dot{\omega}$ cannot be used directly for a
high-precision mass determination, due to the large contribution of frame
dragging.\label{fig:mass}}
\end{center}
\end{figure}

In this Subsection we present the simulations for the mass determination. For
this we assumed five years of observations with weekly TOAs which contain white
Gaussian noise with a standard deviation of 100\,$\mu$s. Fig.~\ref{fig:mass}
shows the results of our simulations for a typical system configuration.  If
this is not the case then, as outlined above, $\dot\omega$ cannot a priori be
used for a high precision mass measurement due to an unknown contribution from
the frame dragging, as we will show later.

In practice, not only one single relativistic effect will be used to determine
the mass of \sgr, but a consistent model, accounting simultaneously for frame
dragging effects in the orbital motion and the signal propagation, will be used
to determine the mass and spin at the best level. How the spin of \sgr\ affects
the timing observations and how it can be extracted from the timing data is the
subject of the next Section.

%%%%%%%%%%%%%%%%%%%%%%%%%%%%%%%%%%%%%%%%%%%%%%%%%%%%%%%%%%%%%%%%%%%%%%%%%%%%%%%%

\section{Frame dragging, spin measurement, and GR's cosmic censorship
conjecture} \label{sec:spin}

Although there are clear indication that the \sgr\ rotates, its actual rate of
rotation is still not well determined. Investigations of flares from accreting
gas in the near-infrared and in X-rays yield a range of $\chi \approx 0.22$ to
$0.99$ \citep{gso+03,agpp04,btdg+06,asc10}. The rather large range in the
estimates of $\chi$ is also a result of the uncertainty in the underlying model
assumptions. A pulsar, however, would provide a clean probe of the gravitational
field of \sgr\ and, in absence of any major external perturbations, give a
direct access to the dragging of inertial frames in the vicinity of \sgr. In
\cite{wk99} it has been shown that in relativistic pulsar--black hole binaries
the (additional) precession of a pulsar orbit due to the frame dragging caused
by the spin of the black hole (Lense-Thirring precession) is the most promising
effect to determine the direction and magnitude of the black hole spin. This, in
general, is also the case for a pulsar in orbit about \sgr. The assumption made
in \cite{wk99}, that the spin of the black hole $S$ is clearly smaller than the
orbital angular momentum $L$, is no longer valid here. The fraction between the
spin of the black hole and the angular orbital momentum is given by
\begin{equation}
  \frac{S_{\rm BH}}{L} = \frac{M_{\rm BH}}{M_{\rm PSR}}\,\beta_e\,\chi \;,
\end{equation}
where $\beta_e = \beta_{\rm O} / \sqrt{1 - e^2}$ and $M_{\rm PSR}$ is the mass
of the pulsar. For a pulsar with an orbital period of less than one year $S_{\rm
BH}/L$ is greater than $50 000 \,\chi$. Thus, the total angular momentum ${\bf
J}$ of the system is completely dominated by the spin of \sgr, whose direction
will therefore practically coincide with the direction of ${\bf J}$, and can,
for the considerations here, be viewed as a constant in time. In this case, the
orbital motion to post-Newtonian accuracy including first-order spin terms can
be taken from Appendix B in \cite{wex95}. This is sufficient to simulate all the
relevant effects \citep[see][for higher order corrections]{kg05} for a
system free of external perturbations. It also accounts for the fact that the
precession is stronger near the pericenter.

%%%%%%%%%%

\subsection{Spin determination from the timing parameters}
\label{ssec:spin deter}

\begin{figure}[ht]
\begin{center}
\includegraphics[width=8.5cm]{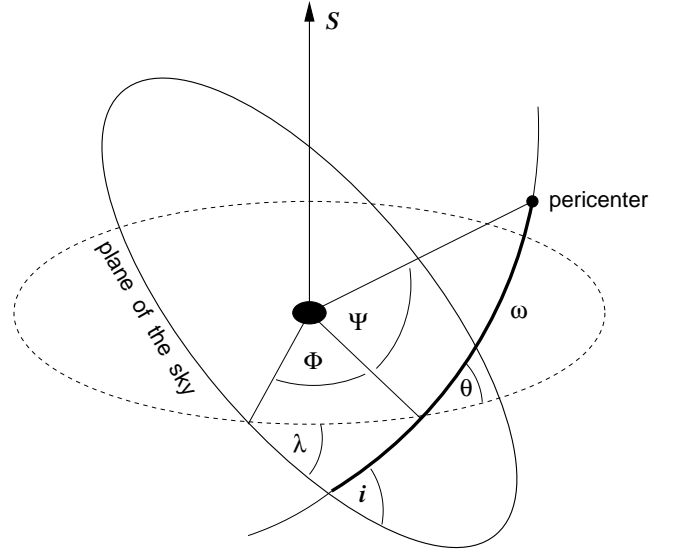}
\caption{Definition of angles in \sgr\ spin reference frame. The
orientation of the orbit with respect to the observer is given by
the orbital inclination $i$ and the longitude of pericenter $\omega$
as measured from the ascending node in the plane of the sky. The
pulsar orbit with respect to the equatorial plane of the rotating
black hole is determined by the inclination $\theta$, the equatorial
longitude of the ascending node $\Phi$, and the equatorial longitude
of pericenter $\Psi$. The angle between the line-of-sight and the
\sgr\ spin is denoted by $\lambda$. \label{fig:angles}}
\end{center}
\end{figure}

Averaging over one orbit, one obtains the rates of the secular precession of the
pulsar orbit caused by frame dragging \citep{bo75}\footnote{To estimate
the measurability of the Lense-Thirring effect, it is sufficient to use the
averaged precession rate. In practice, the preccession of the orbital plane is
more complicated as can be seen from the analytic solution given in Appendix B
of \cite{wex95}.}:
\begin{equation}\label{eq:PhidotPsidot}
\left.
\begin{array}{lcl}
  \dot\Phi &=& \Omega_{\rm LT} \\
  \dot\Psi &=& -3\,\Omega_{\rm LT}\,\cos\theta
\end{array}
\right\} \quad
  \Omega_{\rm LT} = \frac{4\pi}{P_{\rm b}}\,\beta_e^3\,\chi
                  \equiv \hat{\Omega}\,\chi \;. \label{eq:Phi&Psidot}
\end{equation}
The definitions of the angles $\Phi$, $\Psi$ and $\theta$ are given in
Fig.~\ref{fig:angles}. The secular changes for the angles $\Phi$ and
$\Psi$ are linear in time. As discussed in detail in \cite{wex98} and
\cite{wk99}, this linear-in-time evolution translates into a non-linear-in-time
evolution of the observable angles that enter the timing model for a pulsar,
i.e.~the longitude of pericenter $\omega$ and the inclination of the orbit with
respect to the line-of-sight $i$ (as part of the projected semi-major axis $x$).
One finds:\footnote{We define $c_X \equiv \cos X$ and $s_X \equiv \sin X$.}
\begin{equation}\label{eq:ci}
  c_i = c_\theta c_\lambda - s_\theta s_\lambda c_\Phi
\end{equation}
and
\begin{equation}\label{eq:omPsi}
  \sin(\omega - \Psi) = \frac{s_\lambda \, s_\Phi}{s_i} \;, \quad
  \cos(\omega - \Psi) = \frac{c_\lambda - c_\theta\,c_i}{s_\theta\,s_i} \;.
\end{equation}
Since the angles $i$, $\theta$, and $\lambda$ are in the range $0$ to $\pi$,
their sines $s_X$ are non-negative and can be expressed as $s_X = \sqrt{1 -
c_X^2}$. As shown by \cite{wex98}, if the change in $\Phi$ is small (less
than a few degrees) over the time span of the timing observations, the most
straightforward way to analyze the timing data is to fit for the coefficients of
the Taylor expansion of $\omega(t)$ and $x(t)$
\begin{eqnarray}
  \omega &=& \omega_0 + \dot\omega_0(t - t_0)
                      + \mfrac{1}{2}\ddot\omega_0(t - t_0)^2 + \ldots\;,\\
  x      &=& x_0      + \dot x_0    (t - t_0)
                      + \mfrac{1}{2}\ddot x_0    (t - t_0)^2 + \ldots\;,
\end{eqnarray}
and to use the parameters $\omega_0$, $x_0$ and their time derivatives as
intermediate parameters to determine the angles $\theta$, $\lambda$, $\Phi_0$,
$\Psi_0$, and the spin parameter $\chi$. For the configurations considered
in this paper it is sufficient to keep only terms up to second order in $t-t_0$.
Nevertheless, we have extended TEMPO to account for cubic terms in oder to test
their significance in all our simulations. We would like to note in passing,
that the coefficients of the cubic terms can be calculated from the other
coefficients based on basic geometric relations, and therefore they would not
add further information concerning the orientation of the system and the spin
magnitude.

From the derivatives of Eqs.~(\ref{eq:ci}) and (\ref{eq:omPsi}) one finds the
relation between the time derivatives, the orientation of the orbit at a given
epoch, and the spin of \sgr. In practice, the linear trend becomes visible in
the timing data soon after the beginning time of observation, allowing the
measurement of $\dot{x}_0$ and the extraction of the Lense-Thirring contribution
from $\dot\omega_0$. One finds (for convenience we drop the index 0):
\begin{eqnarray}
  \dot{x}  &=&  -x s_i^{-2} c_i s_3 \Omega_{\rm LT} \;,
    \label{eq:xdot} \\
  \dot{\omega} - \dot{\omega}_M &=&
    s_i^{-2}\left[(1 - 3s_i^2)c_\theta - c_i c_\lambda\right] \Omega_{\rm LT}
    \;, \label{eq:omdot}
\end{eqnarray}
where $s_3 \equiv s_\theta s_\lambda s_\Phi$. Since at this point $x$, $|c_i|$,
and $s_i$ are known quantities, the measurement of $\dot{x}$ determines the
quantity $|s_3|\chi$ which must not exceed unity, since according to the CCC
$\chi \le 1$ and $|s_3| \le 1$ by definition. Hence, as soon as $\dot{x}$
becomes measurable one has a first test for the CCC.

To fully determine the magnitude and orientation of the spin, the measurement of
higher order derivatives is necessary. The second time derivatives read
\begin{eqnarray}
  \ddot{x} &=&  -x s_i^{-4} \left[s_3^2 + s_i^2 c_i (c_\theta c_\lambda - c_i)
    \right] \Omega_{\rm LT}^2 \;, \label{eq:x2dot}\\
  \ddot{\omega} &=& s_i^{-4}
    \left[2 c_i c_\theta - (2 - s_i^2) c_\lambda\right]
    \, s_3 \Omega_{\rm LT}^2 \;, \label{eq:om2dot}
\end{eqnarray}
which give us now, in total, six equations (\ref{eq:ci}, \ref{eq:omPsi},
\ref{eq:xdot}, \ref{eq:omdot}, \ref{eq:x2dot}, \ref{eq:om2dot}) for five
unknowns ($\theta$, $\lambda$, $\Phi_0$, $\Psi_0$, $\chi$). For a discussion of
the solution of these equations, we introduce the variables $\chi_\theta \equiv
c_\theta\chi$, $\chi_\lambda \equiv c_\lambda\chi$, and $\zeta_3 \equiv
s_3\chi$. The parameters $\chi_\theta$ and $\chi_\lambda$ represent the
projection of the (normalized) spin onto the orbital angular momentum and the
line-of-sight direction, respectively. They can be determined from the timing
parameters via the equations of the first time derivatives
\begin{eqnarray}
   -\frac{\dot{x}s_i^2}{x\hat{\Omega}} \equiv
     {\cal X}_1 &=& c_i \zeta_3 \;,
     \label{eq:dx_xi}\\
   \frac{(\dot{\omega} - \dot{\omega}_M)s_i^2}{\hat{\Omega}} \equiv
     {\cal W}_1 &=& (1 - 3s_i^2)\chi_\theta - c_i\chi_\lambda \;,
     \label{eq:dom_xi}
\end{eqnarray}
and those of the second time derivatives
\begin{eqnarray}
   \frac{(\ddot{x}x c_i^2+\dot{x}^2 s_i^2)s_i^4}{(x\hat{\Omega}c_i)^2}
   \equiv {\cal X}_2 &=& c_i^2(\chi_\theta^2 + \chi_\lambda^2)
          - c_i(1+c_i^2)\chi_\theta\chi_\lambda \;,\label{eq:ddx_xi}
   \nonumber\\&&\\
   -\frac{\ddot{\omega}x c_i^2s_i^2}{\dot{x}\hat{\Omega}} \equiv
     {\cal W}_2 &=& 2c_i^2\chi_\theta -c_i(1+c_i^2)\chi_\lambda
     \label{eq:ddom_xi}\;.
\end{eqnarray}
where $\zeta_3$ has been eliminated using Eq.~(\ref{eq:dx_xi}). The quantities
${\cal X}_1$, ${\cal W}_1$, ${\cal X}_2$, and ${\cal W}_2$ are defined such that
they do not change when the sign of $c_i$ is flipped. The above equations can be
easily solved analytically. By the time the second derivatives are observable,
$s_i$, $\dot{x}$, $\dot{\omega}_{\rm LT}$, and so ${\cal X}_1$ and ${\cal W}_1$,
will be known with high precision. For a given sign of $c_i = \pm \sqrt{1 -
s_i^2}$, Eq.~(\ref{eq:ddom_xi}) will lead to a unique solution for
$\chi_\theta$, $\chi_\lambda$, and $\zeta_3$. For some orientations
Eq.~(\ref{eq:ddx_xi}) turns out to be more constraining. However, this gives us
in general two solutions for $(\chi_\theta,\chi_\lambda)$. But then
Eq.~(\ref{eq:ddom_xi}), although less constraining, can be used to rule out one
of the two solutions. The best way to represent the solution is to plot the
constraints from Eqs.~(\ref{eq:dom_xi}), (\ref{eq:ddx_xi}) and
(\ref{eq:ddom_xi}) in the $\chi_\theta$--$\chi_\lambda$ plane. Possible
solutions are represented by the region where all three curves meet within the
uncertainty given by the measurement errors of $\dot{x}$, $\dot\omega_{\rm LT}$,
$\ddot{x}$, and $\ddot{\omega}$. This will become clear in the Section below,
where we present the simulations.

With $\chi_\theta$, $\chi_\lambda$ and $\zeta_3$ known, we can calculate the
spin parameter of the black hole via
\begin{equation}\label{eq:chi}
  \chi = s_i^{-1}\sqrt{\zeta_3^2 + \chi_\theta^2 + \chi_\lambda^2
           -2 c_i \chi_\theta \chi_\lambda} \;.
\end{equation}
Once $\chi$ is determined, we can calculate all the angles. Finally,
the $i \leftrightarrow \pi - i$ ambiguity leaves us with two
different solutions in the orientation by $(\Phi_0,\lambda)
\leftrightarrow (\pi+\Phi_0, \pi-\lambda)$, for which $\chi$ has the
same value.

%%%%%%%%%%%%%%%%%%%%%%%%%%%%%%%%%%%%%%%%%%

\subsection{Simulations} \label{sec:spin_simu}

\begin{figure}[ht]
\begin{center}
\includegraphics[width=8.5cm]{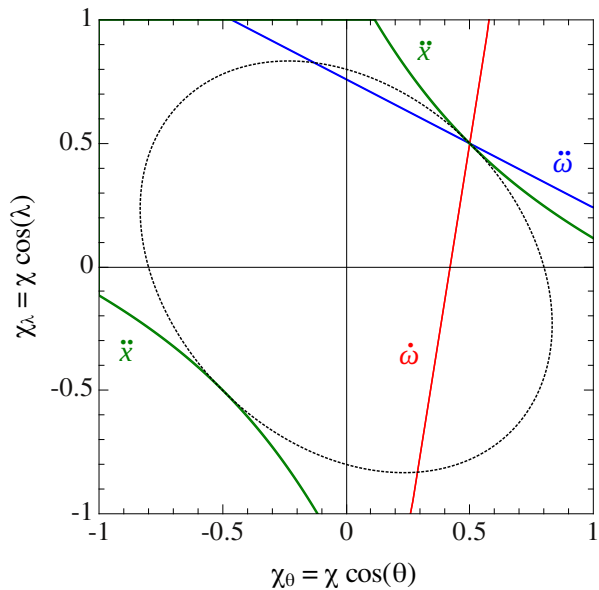}
\caption{Determination of the \sgr\ orientation in the
$\chi_\theta$--$\chi_\lambda$ plane. For this simulation we have
used an orbital period of 0.3\,yr, an orbital eccentricity of 0.5,
$\chi = 1$, $\Psi_0 = 45^\circ$, $\Phi_0 = 45^\circ$, $\theta =
60^\circ$, and $\lambda = 60^\circ$, which are in agreement with the
constraints by \cite{zed+11}. A change in the sign of $c_i$ mirrors
the figure along the $\chi_\lambda=0$ line, meaning that the
solution for $\theta$ is invariant, but $\lambda$ changes to $\pi -
\lambda$. The corresponding spin parameter, as calculated from
Eq.~(\ref{eq:chi}), is $\chi = 0.9997 \pm 0.0010$ (95\% C.L.). In
all the $\chi_\theta$--$\chi_\lambda$ plots (Fig.~\ref{fig:spinsol1}
-- \ref{fig:spinsolfail}) we plot the 68\% confidence intervals.
However, in most cases the separation between the two lines is below
the resolution of the plot. The dotted ellipse is the boundary of
the area for Kerr black holes (see text for details).
\label{fig:spinsol1}}
\end{center}
\end{figure}

\begin{figure}[ht]
\begin{center}
\includegraphics[width=8.5cm]{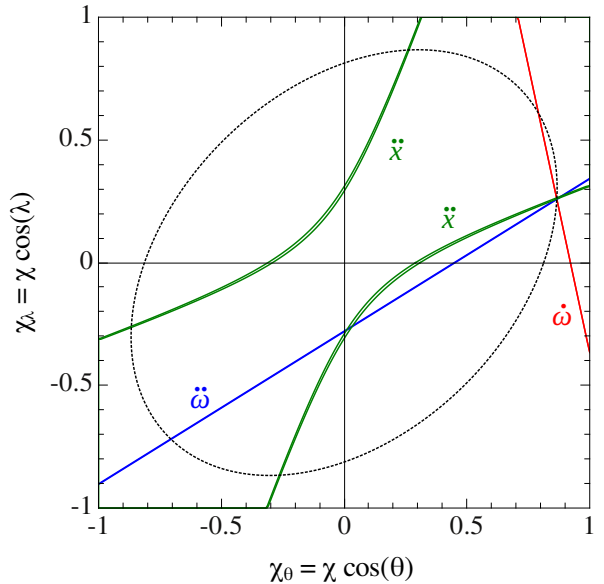}
\caption{Like Fig.~\ref{fig:spinsol1}, but $\Phi_0 = 105^\circ$,
$\theta = 30^\circ$, and $\lambda = 75^\circ$. The corresponding
spin parameter, as calculated from Eq.~(\ref{eq:chi}), is $\chi =
1.0001 \pm 0.0003$ (95\% C.L.). \label{fig:spinsol2}}
\end{center}
\end{figure}

The technique described in the previous Subsection has been tested by a
set of standard simulations for various orbital configurations. For a given
system, following the procedures described in Section~\ref{ssec:mass_simu} we
simulate weekly 100\,$\mu$s TOAs over a time span of five years. Here in the
calculations of the time delays, in addition to the relativistic pericenter
advance we also consider the influence of \sgr\ spin by inputting the secular
changes of $\Phi$ and $\Psi$ described in Eq.~(\ref{eq:Phi&Psidot}). Then, in
order to determine the PK parameters, we fit the TOAs with the MSS timing model
of TEMPO, which we have extended to model the secular changes in pericenter and
projected semi-major axis up to the third order in the time derivatives. The
third order coefficient turn out not to be significant in the simulations
presented here. Figs.~\ref{fig:spinsol1} and \ref{fig:spinsol2} show the
$\chi_\theta$--$\chi_\lambda$ plane for two different orientations of the black
hole and the pulsar orbit. According to GR the solution has to lie within the
boundaries of the figures, since $-1 \le \chi_\theta,\chi_\lambda \le 1$ for a
Kerr black hole. Moreover, the solution ($\chi_\theta, \chi_\lambda$) has to lie
within an ellipse defined by setting $\chi = 1$ in Eq.~(\ref{eq:chi}), in order
to represent a Kerr black hole with an event horizon. Once $\dot{x}$ is
measured, one can determine $\zeta_3$ from Eq.~(\ref{eq:dx_xi}) and use this
quantity to plot the ellipse defined by Eq.~(\ref{eq:chi}) in the
$\chi_\theta$--$\chi_\lambda$ plane.

\begin{figure}[ht]
\begin{center}
\includegraphics[width=8.5cm]{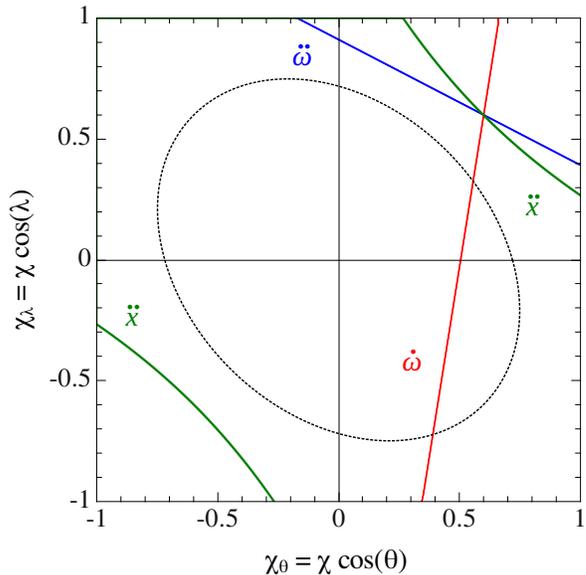}
\caption{Parameters as in Fig.~\ref{fig:spinsol1}, but $\chi=1.2$
(naked Kerr singularity). The dotted ellipse is the (outer) border
of the region where, for the measured orbital inclination and
$\dot{x}$, the Kerr black holes are located, i.e.~where $\chi \le
1$. \label{fig:spin_ns}}
\end{center}
\end{figure}

Fig.~\ref{fig:spin_ns} shows a simulation for a Kerr solution with a spin that
exceeds the spin of an extreme Kerr black hole. Within GR, this would represent
a naked singularity. For such an object the CCC is violated and the
predictability of the (classical) theory breaks down. Also, all three lines have
to agree in a common region, otherwise either GR is not the correct theory, or
there are external perturbations present, a situation which we discuss in more
details in the next Section.

%%%%%%%%%

\subsection{Identification of external perturbations}
\label{ssec:perturb_spin}

\begin{figure}[ht]
\begin{center}
\includegraphics[width=8.5cm]{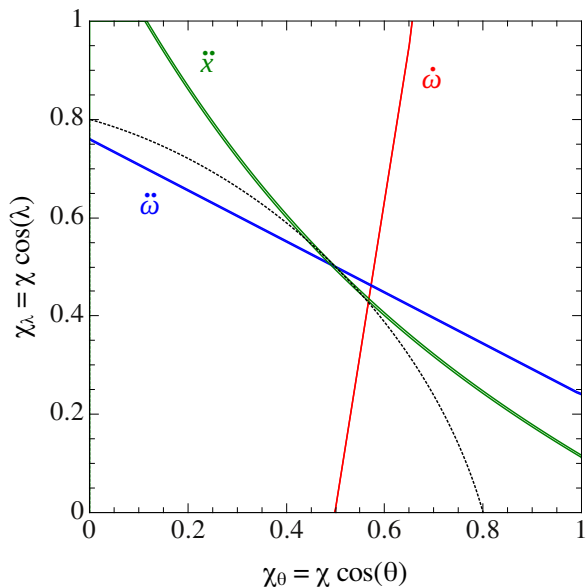}
\caption{Parameters as in Fig.~\ref{fig:spinsol1}, but the
precession of pericenter has an additional contribution from an
external perturbation that amounts to 10\% of the Lense-Thirring
contribution. For a better resolution only the first quadrant of
Fig.~\ref{fig:spinsol1} is plotted here. \label{fig:spinsolfail}}
\end{center}
\end{figure}

As discussed in great detail by \cite{mamw10}, the orbit of a star or pulsar
around \sgr\ may be subject to perturbations from other stars in the vicinity of
the black hole. Depending on the number density of the stars, this could
significantly affect the precession of the pulsar orbit. Nevertheless, since we
have three lines in the $\chi_\theta$--$\chi_\lambda$ plane that need to
intersect, our analysis will unveil the presence of any external perturbations.
In Fig.~\ref{fig:spinsolfail} we present a $\chi_\theta$--$\chi_\lambda$ diagram
based on timing data that contain (besides the gravitational field of \sgr) an
external perturbation causing an additional precession of the pericenter. For
orbits with $P_{\rm b} \la 0.3\,{\rm yr}$, even a small (compared to the
Lense-Thirring precession) external contribution to the precession of the
pericenter leads to a situation where the $\dot\omega$, $\ddot\omega$, and
$\ddot{x}$ lines fail to intersect in one point within the measurement
precision. The same is true, if there is an external contribution to a change in
the inclination of the orbital plane. Hence, if all three lines intersect, we
not only have a precise determination of the spin of the black hole, but also a
test that this measurement is not contaminated by external perturbations.

At this point we would like to add a more detailed comment on the
discriminating power of the pulsar test concerning external perturbations. In
practice, the three-line test outlined above is not simply based on the secular
precession rates. We emphasize that a consistent fit of the timing data, with a
model that includes the Lense-Thirring precession, needs to incorporate the full
dynamics of the orbital precession as given by Appendix B in \cite{wex95}. The
phase dependence of the Lense-Thirring precession rate is a direct result of the
Coriolis type force caused by the dipolar gravitomagnetic field of the central
rotating black hole. Hence, we can identify an external perturbation based on
this quasi-periodic effect, even in a fine tuned situation where the external
mass distribution manages to mimic a secular Lense-Thirring precession. In fact,
we have conducted simulations and found that the phase dependent precession rate
leads to an effect that is typically four orders of magnitude larger than the
timing precision assumed in our simulations. This is in line with the findings
of \cite{dd86}, who pointed out the strength of quasi-periodic effects in tests
of gravity.

%%%%%%%%%%%%%%%%%%%%%%%%%%%%%%%%%%%%%%%%%%%%%%%%%%%%%%%%%%%%%%%%%%%%%%%%%%%%%%%%

\section{Quadrupole measurement and no-hair theorem test}
\label{sec:quadrupole}

The quadrupole moment of the black hole leads to an additional secular
precession of the pulsar orbit. This precession, however, is even for compact
orbits ($P_b \sim 0.1$ yr) much smaller than the Lense-Thirring precession.
Further, it can be shown that the secular terms of the precession cannot be
separated from the Lense-Thirring effect. For this reason, it has been argued by
\cite{wk99} that while the spin magnitude and the orientation of the black hole
are mainly determined by the overall precession of the orbit, the quadrupole of
the black hole is mostly determined via its periodic influence on the motion of
the pulsar from one pericenter to the next. As will be shown in this Section,
these periodic features of the quadrupole can be used to fit for the quadrupole
moment of \sgr.

%%%%%%%%%%%

\subsection{Extracting the quadrupole from the timing data}
\label{ssec:q_motion}

\begin{figure}[ht]
\begin{center}
\includegraphics[width=8.5cm]{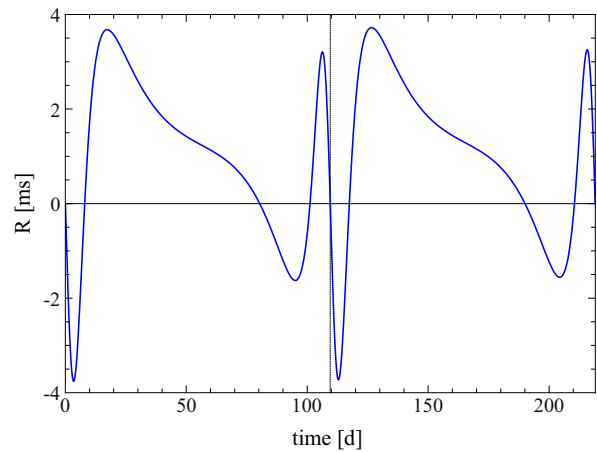}
\caption{Residuals caused by the quadrupole moment of \sgr\ plotted
for two orbital phases. We have used the same orbital and black hole
parameters as in Fig.~\ref{fig:spinsol1}. \label{fig:qres}}
\end{center}
\end{figure}

The deviations in the motion of the pulsar caused by the quadrupole
moment lead to a variation in the Roemer delay, which we describe by
a change in the coordinate position of the pulsar according to
\begin{equation}
  {\bf r}' = (r + \delta r^{(q)})(\hat{\bf n} + \delta\hat{\bf n}^{(q)})
  \;. \label{eq:qmotion}
\end{equation}
The vector $\delta\hat{\bf n}$ is calculated from the changes in the
angles
\begin{equation}
  \Phi'   = \Phi   + \delta\Phi^{(q)}   \;, \quad
  \Psi'   = \Phi   + \delta\Psi^{(q)}   \;, \quad
  \theta' = \theta + \delta\theta^{(q)} \;,
\end{equation}
according to $\delta\hat{\bf n} = \hat{\bf n}' - \hat{\bf n}$. To first order in
$\epsilon \equiv -3Q/a^2(1-e^2)^2$, the detailed equations for the
$\delta$-quantities can be taken from \cite{g59}, with slight modifications that
account for the dominating precession of pericenter caused by the mass monopole:
the term $(5y^2 - 1)$ in the auxiliary constants $m$ and $\gamma$ has to be
replaced by $2\dot\omega P_{\rm b}/\pi\epsilon$, where $\dot\omega$ is the total
advance of pericenter. Based on this we have developed a timing model that
includes the contribution of the quadrupole moment of \sgr\ to first order in
$\epsilon$. Fig.~\ref{fig:qres} illustrates the unique periodic timing residuals
caused by the quadrupole moment of \sgr.

This periodic signal will not only allow the determination of the quadrupole
moment of \sgr\ with high precision, but also provide a clear identification of
the quadrupolar nature of the gravitational field. Moreover, due to the large
advance of pericenter the quadrupolar signal will change in a characteristic way
from one orbit to the next. This clearly helps to identify any external
``contamination'' of the orbital motion of the pulsar, and, as in the spin
determination, provides high confidence in the reliability of a no-hair theorem
test with a pulsar around \sgr.

%%%%%%%%%%%%%%%%%%%%%%%%%%%%%%%%%%%%%%%%%%%%%%%%%%%%%%%

\subsection{Simulations} \label{ssec:q_simu}

\begin{figure}[ht]
\begin{center}
\includegraphics[width=8.5cm]{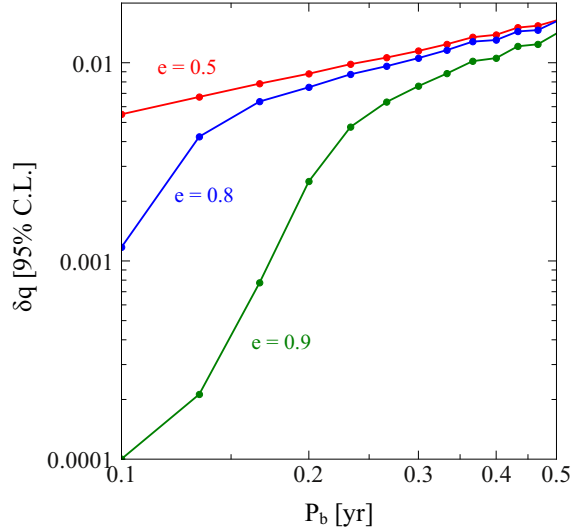}
\caption{Measurement precision for the quadrupole moment of \sgr\ as
a function of orbital period for three different eccentricities, in
absence of any external perturbations. We have used the same orbital
and black hole parameters as in Fig.~\ref{fig:spinsol1}. For the
timing, we assumed the same time span and characteristics of TOAs as
in Fig.~\ref{fig:spinsol1}. This time however the TOAs were equally
distributed with respect to the true anomaly, in order to account
for the fact that timing needs to be done more frequently around
pericenter to optimize the measurement of the quadrupolar signal in
the TOAs. \label{fig:qerr}}
\end{center}
\end{figure}

We have tested the procedure outlined above in a number of mock data
simulations, for various orbital configurations. Again following the
procedures described in Section~\ref{ssec:mass_simu} we assume weekly TOAs with
a precision of 100\,$\mu$s for a time span of five years. This time we
extended our simulations and the timing model used in
Section~\ref{sec:spin_simu} to account for the periodic effects due to
quadrupole moment of \sgr\ described in Eq.~(\ref{eq:qmotion}). Our results are
summarized in Fig.~\ref{fig:qerr}. Note that the precision of the spin
determination is at least one order of magnitude better than the determination
of $q$. Hence the uncertainty in the $q$-measurement is the limiting factor for
the no-hair theorem test. As a conclusion of our simulations, if the
external perturbations are negligible, for orbits with $P_{\rm b} \lesssim
0.5$\,yr the no-hair theorem can be tested with high precision. If we adopt the
precessional rates from stellar perturbation calculated in
Fig.~\ref{fig:precession_rates}, we conclude that the test can be achieved with
high precision for orbits with $P_{\rm b}\lesssim 0.1$\,yr. This range can be
extended if in the presence of perturbations the characteristic quadrupolar
features remain separable. This, however depends on the details of the external
mass distribution, which we will not investigate further in this paper.

%%%%%%%%%%%%%%%%%%%%%%%%%%%%%%%%%%%%%%%%%%%%%%%%%%%%%%%%%%%%%%%%%%%%%%%%%%%%%%%%

\section{Discussion}
\label{sec:discussion}

In this paper we have developed a method to determine the mass, the spin, and
the quadrupole moment of \sgr\ using a pulsar in a compact orbit around this
super-massive black hole. Our investigation is based on a consistent timing
model, that includes all the relativistic and precessional effects that can be
used to extract these parameters of \sgr. Based on simulated timing data for a
pulsar in orbit around \sgr, we have shown in a consistent covariance analysis,
that even with a moderate timing precision ($\sim 100\,\mu{\rm s}$), one can
expect to be able to determine the mass, the spin, and the quadrupole moment of
\sgr\ with high precision, provided the orbital period is well below one year.
As a result of our simulations, for a compact orbit (orbital period of a few
months) one can expect to measure the spin with a precision of $10^{-3}$, or
even better. We have shown how the method would allow the identification of an
object whose frame-dragging exceeds that of an extreme Kerr black hole, and
therefore would provide a test of the CCC. Furthermore, for such orbits the
determination of the quadrupole moment of \sgr\ seems feasible at a few percent
precision level or even better, depending on the size and orientation of the
pulsar orbit and the spin of \sgr. In combination with the precise spin
measurement from the Lense-Thirring effect, this yields a high precision test of
the no-hair theorem of stationary black holes.

Moreover, we have shown that, in general, our analysis will be able to unveil
the presence of external perturbations caused by the presence of distributed
mass, therefore providing high confidence in a spin and quadrupole determination
based on pulsar timing. Nevertheless, further studies are required to see
whether the spin and quadrupole moment can still be extracted if the timing data
is ``contaminated'' by external perturbations. If perturbations arise from a
smooth concentration of dark matter particles in the vicinity of \sgr, we may be
able to learn something about the properties of dark matter that clusters around
\sgr, assuming GR is correct.

Finally, we need to emphasize that the tests presented are not affected by an
uncertainty in the distance to the GC. On the contrary, a mass determination via
pulsar timing would give a greatly improved value for $R_0$ if combined with the
astrometric measurements in the near infrared.

Once a pulsar is detected in a compact orbit around \sgr, continuous timing will
allow more and more measurements and tests as the timing baseline grows with
time. In the following we summarize the most important steps in this experiment:
\begin{itemize}
\item After timing one orbit, all Keplerian parameters will be well known,
      and also the pericenter advance will be measured with good
      precision. This will already provide a good estimate of the mass of \sgr.
\item Timing a few more orbits would then allow the determination of additional
      post-Keplerian parameters, like the Shapiro parameters ($r_{\rm Sh}$,
      $\sin i$) and the amplitude of the Einstein delay ($\gamma_{\rm E}$).
      These parameters allow a robust determination of \sgr\ mass and the
      inclination of the pulsar orbit with respect to the line-of-sight.
\item At this stage, a measurement of a change in the projected semi-major axis
      ($\dot{x}$) caused by the Lense-Thirring will allow an early test of
      the CCC, and mapping of the region in the $\chi_\theta-\chi_\lambda$
      plane where the solutions for Kerr black holes are.
\item Around the same time the mass measurement should reach a precision
      that allows the extraction of the Lense-Thirring contribution to the
      precession of the pericenter ($\dot{\omega}_{\rm LT}$), giving a
      line-like region in the $\chi_\theta-\chi_\lambda$ plane.
\item After a few years of timing the second time derivatives of
      $\omega$ and $x$ should be known with high precision, allowing a
      precise determination of the \sgr\ spin (magnitude and direction).
      At this stage we also have a test for the ``cleanness'' of the
      system, and whether the spin is below the Kerr bound ($\chi = 1$).
\item At the same time the obtained parameters for the pulsar orbit and the
      \sgr\ spin can be used to model the periodic features in the timing
      residuals, which are caused by the quadrupole moment of \sgr. This
      allows to a determination of the quadrupole moment and a test of the
      no-hair theorem.
\end{itemize}

A potential problem for the timing of a pulsar in a compact orbit around \sgr\
is posed by the relativistic spin precession, as pointed out by \cite{mamw10}.
This change in the pulsar orientation with respect to a distant observer not
only causes a variation of the pulse profile, which makes precise timing more
difficult, but also can turn the pulsar emission away from our line-of-sight
\citep{wrt89,kra98}. To leading order the spin-precession is given by the de
Sitter precession rate, which for $M_{\rm BH} \gg M_{\rm PSR}$ reads
\citep{bo75b}
\begin{equation}
  \Omega_{\rm dS} \simeq \frac{3\pi}{P_{\rm b}}\,\beta_e^2
                  \simeq (0.13^{\circ}{\rm /yr})\,\frac{1}{1 - e^2}
                         \left(\frac{P_{\rm b}}{1\,{\rm yr}}\right)^{-5/3}
                          \;.
\end{equation}
Consequently, for orbital periods below one year relativistic spin
precession is expected to play an important role in the timing
observations.  We note in passing, that the Pugh-Schiff precession
rate caused by frame dragging \citep[$\Omega_{\rm FD} \sim  2\pi
\beta_e^3\chi/P_{\rm b}$,][]{pug59,sch60} is only relevant in the
case of very compact, highly eccentric orbits ($\Omega_{\rm FD} \sim
1^{\circ}$/yr for $P_{\rm b} = 0.1$\,yr, $e=0.8$ and $\chi=1$), and
could provide an independent test of the rotation of \sgr.

The no-hair theorem test can also be affected by the accretion disc
around \sgr. To get an idea of the strength of this effect, one can
estimate the influence by calculating the fraction of the
quadrupolar potential of the disc to that of the black hole. This
ratio is given by
\begin{equation}
\mathscr{R}\sim M_{\rm disc} r^2_{\rm disc}/(M_{\rm BH}r^2_{\rm
g})\;,
\end{equation}
where $M_{\rm disc}$ and $r_{\rm disc}$ are the mass and outer radius of the
disc, and $r_{\rm g} \equiv GM_{\rm BH}/c^2$ is the black hole gravitational
radius. Following the advection-dominated accretion flow (ADAF) model of
\cite{ych+09} and adopting, as an upper limit, the disc scale of $\approx
1\,{\rm arcsec}$ determined from X-ray observation \citep{bbb+01}, we obtain
$\mathscr{R} \approx 0.4\%$, which indicates that the quadrupole moment
measurement of \sgr\ would not be biased by the contribution of the disc above
the $1\%$ precision level. Furthermore, in a very resent publication,
based on current X-ray and millimeter observations, \cite{psa11} concludes that
for compact orbits, like the ones discussed in this paper, hydrodynamic drag
forces from plasma in the vicinity of \sgr\ are expected to be negligible.

\section*{Acknowledgements}
We are grateful to K.~J.~Lee and G.~X.~Li for valuable discussions, and would
like to thank J.~P.~W.~Verbiest for carefully reading the paper and providing
detailed comments. We also would like to thank the anonymous referee for
his careful review of this manuscript, and for his useful comments. This
research has made use of NASA's Astrophysics Data System Bibliographic
Services. KL is funded by a stipend of the Max-Planck-Institute for Radio
Astronomy.

%%%%%%%%%%%%%%%%%%%%%%%%%%%%%%%%%%%%%%%%%%%%%%%%%%%%%%%%%%%%%%%%%%%%%%%%%%%%%%%%

%%% Bibliography %%%

\bibliographystyle{apj}
\bibliography{journals,psrrefs,modrefs,mycollection,crossrefs}

\end{document}